\documentclass[preprint,showpacs,preprintnumbers,amsmath,amssymb,floats]{revtex4-1}
\usepackage{blindtext}
\usepackage{graphicx}
\usepackage[english]{babel}
\usepackage{amsmath}
\usepackage{amssymb}
\usepackage{color}

\newcommand{\be}{\begin{eqnarray}}
\newcommand{\ee}{\end{eqnarray}}
\usepackage[toc,page]{appendix}

\begin{document}
\title{Phase diagram and quantum-criticality of the two dimensional dissipative quantum XY model}

\author{Changtao Hou and Chandra M. Varma}
\affiliation{Department of Physics, University of California, Riverside, CA}
\date{\today}
\begin{abstract}
 The two-dimensional dissipative quantum XY model is applicable to the quantum-critical properties of diverse experimental systems, ranging from the superconductor to insulator transitions, ferromagnetic and antiferromagnetic transitions in metals, to the loop-current order transition in the cuprates.  We solve the re-expression of this model in terms of topological excitations: vortices and a variety of instantons, by renormalization group methods. The calculations explain the extraordinary properties of the model discovered in Monte-Carlo calculations: the separability of the quantum critical fluctuations (QCF) in space and time, the spatial correlation length proportional to logarithm of the temporal correlation length near the transition from disordered to the fully ordered state, and the occurrence of a phase with spatial order without temporal order. They are intimately related to the flow of the metric of time in relation to the metric of space, i.e. of the dynamical critical exponent $z$. 
These properties appear to be essential in understanding the strange metallic phase found in a variety of quantum-critical transitions as well as the accompanying high temperature superconductivity.
\end{abstract}
\maketitle

The dissipative quantum XY (DQXY) model was introduced \cite{ChakraKivel1} to understand the superconductor to insulator transitions in 2 D films as a function of the
normal state resistance \cite{GoldmanRev2014}. The model is of-course directly applicable to quasi-2D metallic ferromagnets with strong XY anisotropy, a realization of which has been found in the quantum-critical region \cite{AronsonPNAS2014, HouCMV2016}. Quasi-2D metallic anti-ferromagnets, with 
 incommensurate uni-axial order  or commensurate planar order also map to the dissipative XY model \cite{CMV-PRL2015}, while model with incommensurate planar order maps to the closely related $U(1)\times U(1)$ model. Metallic anti-ferromagnets are of great experimental interest; they are realized by the Fe-based compounds, where superconductivity occurs in a region around the antiferromagnetic quantum-critical point, and in several heavy-fermion compounds. The same model also describes the statistical mechanics of the loop-current order proposed for under-doped cuprates \cite{CMV-PRB2006} ending in a quantum-critical point in a region of doping of the highest $T_c$. In the Fe-compounds, in the heavy-fermions and in the cuprates, the normal state singular Fermi-liquid properties in the quantum-critical or strange metal region, for example the entropy, the resistivity and the nuclear relaxation rates have the same singular functional dependence on temperature, despite the complete difference in their microscopic models. This encourages one in seeking a common universality class for their statistical mechanics.
These diverse interesting problems call for a thorough understanding of the phase diagram and the correlation functions of the DQXY model.

It is well known that the classical XY model in 2D does not belong to the Ginzburg-Landau-Wilson (LGW) class of models for phase transitions which in essence are driven by renormalized spin-waves. The classical transition in the 2D-XY model on the other hand is driven by proliferation of vortices  \cite{KT1973, Berezinskii}. The kinetic energy in the pure
quantum XY model turns it into Lorentz-invariant model so that the quantum-transition and the associated critical fluctuations are the same as in the classical 3D XY model near its classical transition. However, on inclusion of dissipation, the model has a much richer phase diagram \cite{Stiansen-PRB2012, ZhuChenCMV2015}. The 2D dissipative quantum XY model (DQXY) can be transformed \cite{Aji-V-qcf1, Aji-V-qcf3} to a model in which the properties are governed by topological excitations, two-dimensional vortices and 'warps'.
Warps are instantons of monopole anti-monopole combinations with zero net charge as well as zero dipole. The order parameter correlation functions discovered by Monte-Carlo calculations \cite{ZhuChenCMV2015} are quite unlike the form expected in extensions of the LGW theories to quantum-critical phenomena \cite{Moriya-book, Hertz}.  In this paper we use the re-expression of the dissipative quantum XY model in terms of warps and vortices and perform renormalization group calculations, which have some interesting new technical aspects, to reproduce the principal features of the phase diagram and of the essential aspects of the correlation functions discovered in the Monte-Carlo calculations. This leads to a deeper understanding of the results obtained by numerical methods.



The action of  the (2+1)D quantum dissipative XY model for the angle $\theta({\bf x}, \tau)$ of fixed-length quantum rotors at space-imaginary time point $({\bf x}, \tau)$ is 
\begin{eqnarray}
\label{action1}
S &=&-K_0 \sum_{\langle {\bf x, x}' \rangle} \int_0^{\beta} d \tau \cos(\theta_{{\bf x}, \tau} - \theta_{{\bf x}', \tau}) 
 + \frac{1}{2E_{0}} \sum_{{\bf x}} \int_0^\beta d \tau \left( \frac{d \theta_{{\bf x}}}{d\tau}\right)^2  \nonumber \\
&+&  \alpha \sum_{\langle{\bf x, x}'\rangle} \int_0^{\beta} d \tau  d\tau' \frac {\pi^2}{\beta^2} \frac {\left[(\theta_{{\bf x}, \tau} - \theta_{{\bf x}', \tau})  -(\theta_{{\bf x}, \tau'} - \theta_{{\bf x}', \tau'}) \right]^2}{
\sin^2\left(\frac {\pi |\tau-\tau'|}{\beta}\right)},
\end{eqnarray}
 $\tau/2\pi$ is periodic in $\beta$, the inverse of temperature $1/(k_B T)$.  $\langle {\bf x, x}'\rangle$ denotes nearest neighbors. The first term is the spatial coupling term as in classical XY model. The second term is the kinetic energy where $E_0$ serves as the moment of inertia. The third term describes quantum dissipations of the ohmic or Caldera-Leggett type~\cite{CaldeiraLeggett}. Such a dissipation also comes from the decay of the current fluctuations of the DQXY model to fermions current-current correlations with resistance per square $R$. In that case $\alpha = 4\pi^2(R_Q/R)$, where $R_Q = h/e^2$ is the quantum of resistance per square.
 
 \begin{figure}[th]
\includegraphics[width=0.6\columnwidth]{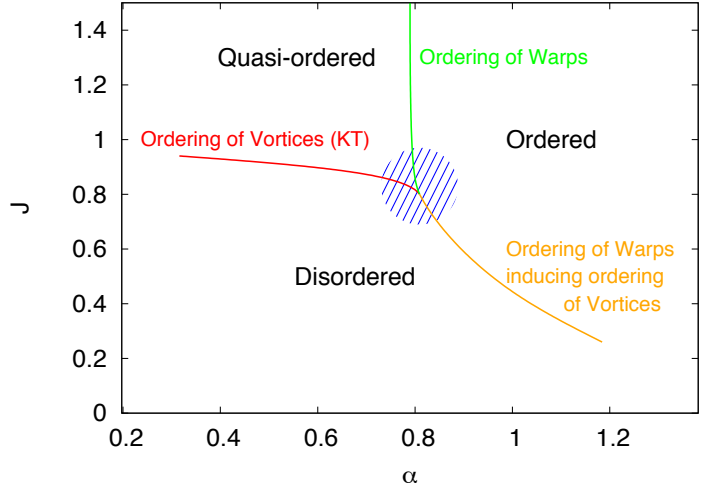}
\caption{Phase diagram for the dissipative 2D-quantum XY Model calculated by Monte-Carlo in \cite{Stiansen-PRB2012, ZhuChenCMV2015}. $J\equiv K_0\tau_c$. The calculations are for a fixed value of the dimensionless variable $E_c \equiv E_0\tau_c) = 100$.  $\tau_c^{-1}$ serves as the ultra-violet cut-off by which $K_0$ and $E_0$ of Eq. (\ref{action1} have been normlaized. More results can be found in \cite{ZhuChenCMV2015, ZhuHouCMV2016}}
\label{fig:phdia}
\end{figure}
 
The phase diagram \cite{Stiansen-PRB2012, ZhuChenCMV2015} for this model is shown in Fig. (\ref{fig:phdia}). There are three lines of transitions, separating the three different phases shown. From the disordered phase, one has a transition by varying the dissipation parameter $\alpha$ or the parameter $\tilde{K} \equiv \sqrt{KK_{\tau}}$ to a phase which has the properties of the ordered  phase of the classical 3D-XY model. By increasing $\tilde{K}$ for small enough $\alpha$, one has a Kosterlitz-Thouless type 2D-vortex induced transition to a quasi-ordered phase, which is spatially ordered but (1D)-disordered in time.  The temporal correlations do not change across this transition. The quasi-ordered phase also orders in time by increasing $\alpha$ to the fully ordered phase. We show how the three phases and the correlation functions at the three lines of transitions come about by a renormalization group analysis, starting with the transformation of the original model to one of purely topological excitations. 
 
 In Refs.\cite{Aji-V-qcf1, Aji-V-qcf3}, it is shown that after making a Villain transformation \cite{Villain} and integrating over the small oscillations or spin-waves, the action is expressed in terms of link variables which are differences of $\theta$'s at nearest neighbor sites,
 \be
\label{eq:m}
m_{{\bf x,x}'}(\tau, \tau') \equiv \theta({\bf x},\tau) - \theta({\bf x}',\tau').
\ee
Quite generally, 
${\bf m} = {\bf m}_{\ell} +{\bf m}_t$, 
where ${\bf m}_{\ell}$, is the longitudinal (or curl-free) part  and ${\bf m}_t$ is the transverse (or divergence-free) part. The appearance of ${\bf m}_{\ell}$ is a novel feature of the quantum dissipative XY-model.
Now define
\be
\nabla \times {\bf m}_t ({\bf x},\tau) = \rho_v({\bf x}, \tau) \hat{\bf z},
\ee
so that $\rho_v({\bf x}, \tau)$ is the charge of the vortex at $({\bf x},\tau)$. The model also has quantized jumps in phase at the point ${\bf x}$ beteen $\tau$ and $\tau +d\tau$. Such jumps produces divergences in 
${\bf m}$ \cite{Aji-V-qcf3} and can be represented by 
 \be
\frac{\partial {\hat{\nabla}}\cdot {\bf m}_{\ell}({\bf x}, \tau)}{\partial \tau} = \rho_w({\bf x}, \tau).
\ee
Although a continuum description is being used for simplicity of writing, it is important to do the calculation so that the discrete nature of the $\rho_v, \rho_w$  fields is always obeyed. 
The action of the model (\ref{action1}) in terms of warps and vortices is \cite{Aji-V-qcf1, Aji-V-qcf3},
\begin{eqnarray}
\label{topomodel}
S &=& \frac{J}{2\pi}\sum_{i\ne j}\rho_v(\bm{r}_i,\tau_i)\ln\frac{|\bm{r}_i-\bm{r}_j|}{a_c}\rho_v(\bm{r}_j,\tau_i) + \alpha \sum_{i\ne j}\rho_w(\bm{r}_i,\tau_i)\ln\frac{|\tau_i-\tau_j|}{\tau_c}\rho_w(\bm{r}_i,\tau_j)\\ \nonumber 
&+&g\sum_{i\ne j}\rho_w(\bm{r}_i,\tau_i)\frac{1}{\sqrt{|\bm{r}_i-\bm{r}_j|^2+v^2(\tau_i-\tau_j)^2}}\rho_w(\bm{r}_j,\tau_j)
 +\ln y_w\sum_{i}|\rho_w(\bm{r}_i,\tau_i)|^2+\ln y_v\sum_{i}|\rho_v(\bm{r}_i,\tau_i)|^2.
\end{eqnarray}
The sum is over all space points  and over imaginary time $\tau$ from an upper cut-off $\tau_c$ to $1/(2\pi T)$. Here $J=K_0\tau_c, g = \sqrt{J/E_c}/4\pi,  v^2/c^2 = J E_c, E_c = E_{0}\tau_c$ are dimensionless variables, and $c = a/\tau_c$, $a$ is the lattice constant. Some spatial dimensions have been absorbed in the re-definition of $\rho_v$ and $\rho_w$ \cite{Aji-V-qcf3}.
The first term in (\ref{topomodel}) is the action of the {\it classical} vortices interacting with each other through logarithmic interactions in space but the interactions are local in time. The second term describes the warps interacting logarithmically in time but locally in space. The third term is the action for a (anisotropic) Coulomb field between warps, which if present alone for the isotropic case is known \cite{polyakov} not to cause a transition; it will be seen to play a crucial role in the present problem in which the space-time anisotropy is required to flow. The short distance core-energy of the warps and vortices is taken care of by the final term in which $y_v$ and $y_w$ are the fugacity of the vortices and the warps, respectively.

 The warp and the vortex variables in the first two terms are orthogonal since they are related respectively to the divergence and rotation of a vector field. The problem is trivial with just these two terms alone. If the first term dominates, one expects a transition of the class of the classical Kosterlitz-Thouless transition through the renormalization of the fugacity of vortices to 0. But the ordered phase would have bound vortex-anti-vortex pairs in space with nothing to correlate them in time. If the second term dominates, there is a quantum transition to a phase with binding of warp-antiwarp pairs in time but nothing to order them with respect to each other in space. Four distinct phases would therefore be found in the $\alpha - JE_{c}$ plane. This is  unlike the phase diagram of Fig. (\ref{fig:phdia}).  We will show that given the growth of correlations due to the renormalization of the density of isolated vortices or of isolated warps $\to 0$, the actual critical points are determined by the third term, which scales time and space differently, depending on whether the warps or the vortices in the first two terms drive the transition. This  leads to ordering at $T=0$ both in time and space to a state with symmetry of the 3D XY model over most of the phase diagram but an interesting region in which the system is spatially ordered for small times but disordered at larger times persists.

The renormalization group (RG) equations for the coupling $J$ and the vortex fugacity $y_v$ may be obtained following the procedure of Kosterlitz \cite{Kosterlitz1974} or Jose et al.\cite{JKKN1977}. The renormalization of these quantities obtained by scaling the spatial length scale ${\ell}_r = \ln (r/a)$, where the lattice constant  $a$ serves as the short-distance cut-off are,
\be
\label{vortrg}
dJ&=&-\pi y_v^2J^{2}d{\ell}_r \\
dy_v&=&(2-\frac{J}{4\pi})y_vd{\ell}_r
\ee
To derive the RG equation for the parameters for the warps, we consider the effective interaction between two warps at a point in space and separated by time $\tau > \tau_c$ as modfied by the screening due to the creation of  a virtual pair, at times $\tau^\prime$ and $\tau^{\prime\prime}$,  $\tau_c<|\tau^\prime-\tau^{\prime\prime}|<\tau_c\text{e}^{d\ell_{\tau}}$, where $\ell_{\tau} = \ln (\tau/\tau_c)$ and $\tau_c$ is the short-time cutoff. We integrate over the coordinates of the two virtual warps to get a renormalized interaction between the real pair. 
The RG equations for $\alpha$ can be derived by scaling $\ell_{\tau}$ in this way. But the fugacity of warps is renormalized by both rescaling $\ell_{\tau}$ and, due to the third term in the action (\ref{topomodel}), by rescaling $\ell_r$. Therefore we must also consider the renormalization of the parameters $g$ and $v$. A scale dependent $v = d|r|/d\tau$ is equivalent to allowing a scale-dependent 
dynamical critical exponent,
\be
\label{z}
z \equiv d{\ell}_{\tau}/d{\ell}_r.
\ee
The renormalization procedure for $y_w$, $g$ and for $v$ are given in a supplementary section. The results are
\be
\label{warp1}
d \alpha &=& -2 \alpha y_w^2 d{\ell}_{\tau} , 
\ee
\be
\label{warp2}
d y_w &=& y_w\big((1-\alpha) d{\ell}_{\tau} + (2-g) d{\ell}_{r}\big), 
\ee
\be
\label{warp3}
d g &=& -gd{\ell}_r-\frac{8\pi^3}{3}\frac{g^2y_w^2}{a_c}\bigg((\frac{1}{4v}+\frac{v}{2})dl_{\tau}+(\frac{1}{v}+\frac{v}{3})dl_r\bigg),
\ee
\be
\label{vz}
d v &=& \big(d{\ell}_{\tau}-d{\ell}_r\big)v.
\ee
These equations may be written as scaling equations either with respect to ${\ell}_r$ or ${\ell}_{\tau}$ by using $z$ defined by Eq. (\ref{z}). For example, (\ref{vz}) may be written as
\be
\label{vz1}
\frac{d v}{d \ell_{\tau}} = (1-z^{-1}) v;
\ee
It is obviously redundant to keep both $z$ and $v$. We note the identity 
\be
\label{identity}
\frac{dz^{-1}}{d{\ell}_{\tau}} - z^{-1}(1- z^{-1})= \frac{\tau}{r} \frac{d v}{d\ell_{\tau}}.
\ee
Using this, (\ref{vz1}) can be re-written as 
\be
\label{eq-z}
\frac{d (z^{-1})}{d l_{\tau}} = 2z^{-1}(1-z^{-1}).
\ee

We now have a closed system of RG equations. 
First, we note that Eq. (\ref{eq-z}) gives the fixed points $z^* = 1, \infty, 0. $ $z^*=1$ is a stable fixed point. From Eq. (\ref{vz1}), we note that near $z^*=1$, the velocity has a stable fixed point at its initial value. The $z^* = \infty$ fixed point is unstable, corresponding to the unstable fixed point for velocity at $v^* = \infty$. The $z^* =0$ fixed point is also unstable, corresponding to  the unstable fixed point at $v^*=0$. These results are in accord with the investigations on expansion about isotropy of the classical anisotropic coulomb gas model in 3D \cite{Kosterlitz1977}, i.e. the model with only the third term in (\ref{topomodel}).
We find that the 2D limit, (i.e. $v^* =0$) as well as the 1D limit ($v^* =\infty$) is unstable (i.e. has a critical point) towards the stable isotropic problem.


We now consider the regimes of initial parameters in which the three different regions in the phase diagram in Fig.(\ref{fig:phdia}) are obtained, and calculate the correlation lengths in time and space about the critical points separating them:

I. $J/2\pi \lesssim 4, \alpha \lesssim 1$: On looking at the first two terms of the transformed action, (\ref{topomodel}), or the RG equations. (\ref{vortrg}, \ref{warp1}, \ref{warp2}), one finds that the fugacity of both vortices and warps is large in this region, provided $g < 2$, as will be shown. So the model is in its quantum disordered state in this region, as in the phase diagram in Fig.(\ref{fig:phdia}).

II. $J/2\pi \lesssim 4$ and $\alpha \approx 1$: In this region, we must first analyze the equations for the warps, Eqs.(\ref{warp1}, \ref{warp2}, \ref{warp3}).
 We note from Eqs. (\ref{warp1}, \ref{warp2}) that for $z^*\to \infty$, and the initial $\alpha >1$, $\alpha$ flows asymptotically for long times to $0$, and $y_w \to 0$, provided $g$ remains finite or zero.  For initial $\alpha <1$, $\alpha$ flows asymptotically at long times to to $0$ and $y_w$ to $\infty$. So $\alpha^*=1$ is an unstable critical point. We note from (\ref{warp3}) that near the $\alpha^*=1$ fixed point, as $z \to \infty$ and $y_w \to 0$, $g$ flows to a constant value, consistent with the above requirement.

We expand near the unstable $\alpha =1^-$ fixed point to find
\be
\label{warpcorr}
y_w \propto \big(e^{-\tau/\xi_{\tau}}-1); ~~ \xi_{\tau} \propto e^{(b_0/(1-\alpha))^{12}}.
\ee
$b_0$ is a coefficient of $O(1)$.
Let us study $J$ and $y_v$ near this point. To do so, we convert all scaling equations in terms of ${\ell}_{\tau}$ by using Eqs. (\ref{z},\ref{eq-z}).
The flow of  the vortex parameters $J$ and $y_v$ is now given by 
\be
\label{inducedJv}
\frac{dJ}{dl_\tau}&=&-\frac{1}{z}\pi y_v^2J^{2},\nonumber\\
\frac{dy_v}{dl_\tau}&=&\frac{1}{z}(2-\frac{J}{4\pi})y_v.
\ee
We also have 
\be
\frac{d y_v}{d y_w} = \frac{1}{z(1-\alpha)}(2-\frac{J}{4\pi}) \frac{y_v}{y_w}.
\ee
Near the critical point $z^* \to \infty$, but $(1-\alpha) \to 0$. So we ask which is more important.
From Eq. (\ref{eq-z}), one finds the leading behavior of $1/z = 0 + O(\tau^{-2})$. But $\alpha$ approaches its fixed point of 1 exponentially slowly with $\tau^{-1}$. So the $(1-\alpha)$ term is not important compared to $z$.
If $(2-J/4\pi)$ does not flow, as is found self-consistently, then indeed
\be
y_v \propto y_w^{1/z}, ~ i.e.~ for~ z \to \infty, ~y_v \propto \ln y_w.
\ee
We can get a correlation length in space from the relation,  
$k \propto \omega^{1/z}$.
For $z \to \infty$, this gives that the spatial correlation length $\xi_r$ is proportional to logarithm of the temporal correlation length $\xi_{\tau}$.
We can also get the same result explicitly from 
$d\ell_r = (1/z) d\ell_{\tau}$ and the result that $1/z \propto \tau^{-2}$ near this fixed point.

The same results for the RG flows are also obtained from the numerical solution of the equations near this critical point. The critical point corresponds to the quantum-disordered  to 3D ordered transition in Fig.(\ref{fig:phdia}). The correlation lengths in time and space deduced above have been found in extensive Monte-carlo calculations \cite{ZhuChenCMV2015}. We understand now that the physical region of the conjecture made in 
 \cite{Aji-V-qcf2, ZhuChenCMV2015} that the freezing of warps drives the freezing of vortices. It is that the growing fugacity of warps drives a flow of the space-time metric parameter $z$ so that the fugacity of the vortices, Eq. (\ref{inducedJv}) becomes scale-dependent even for values of $J$ below the value of $8\pi$.
 
III.  $\alpha \lesssim 1$ and $J/4\pi \approx 2$: In this region, it is appropriate to start the analysis with examination of 
 Eqs. (\ref{vortrg}) for flow of $J$ and $y_v$.
 Eqs. (\ref{vortrg}) have the standard KT flow with the KT point $J^* = 8\pi$ near which $y_v \to 0$. For $J > 8\pi$, $y_w$ flows towards $\infty$ and  $J$ flows to $\infty$. Following Nelson and Kosterlitz, the spatial stiffness has a jump at the transition. Now we examine whether this is changed by the action of warps.
 We start by assuming that such a fixed point corresponds to the $z \to 0$ unstable fixed point. We will soon check that this is consistent. From Eq. (\ref{vz1}),  $z \to 0$ leads to $v \to 0$ at the fixed point. Let us study how warps are affected by this.  Eq. (\ref{warp3}) gives that $g$ flows to 0.
 The equations (\ref{warp1}), (\ref{warp2}) should now be written in terms of the scale length $\ell_r$ as 
 \be
 \frac{d \alpha}{d\ell_r} &=& -2 z \alpha y_w^2 \\  
 \frac{d y_w}{d\ell_r} &=& y_w\big(z(1-\alpha) + 2 \big).
 \ee
 We note that neither the fugacity $y_w$ nor $\alpha$ flow in this case. So warps remain completely unaffected by the vortex freezing. The time dependence of the correlation remains unaffected, as can be checked directly. This is consistent with the assumption that this fixed point corresponds to $z^*=0$.
We have a phase in which the  spatial correlations become of the ordered Berezinsky-Kosterlitz-Thouless phase but the correlations in time remain of the disordered phase. This corresponds to the transition from the quantum-disordered phase to the quasi-ordered phase in Fig. (\ref{fig:phdia}). The results are consistent with the Monte-carlo calculations, which give that the transitions at $T \to 0$ in this regime of parameters is a pure Kosterlitz-Thouless transition with a jump of spatial stiffness, with the correlations of the order parameter unchanged from those in the disordered phase. This phase transition may well correspond to the superconductor to a Normal metal transition found in superconducting films \cite{KapitulnikSIT1}. If so, the quasi-ordered phase must be an unusual metal. This matter requires further investigation.
 
IV. $J/2\pi \gtrsim 4, \alpha \to 1^-$: As discussed in III, for these values of $J$, $y_v \to \infty$ and isolated vortices are frozen for $\alpha < 1$. As 
$\alpha \to 1^-$, $y_v$ remains stable at this value and the RG equations for $\alpha$ and $y_w$ are simply (\ref{warp1}) and (\ref{warp2}), respectively. So $y_w \to 0$ as $\alpha \to 1$ and density of isolated warps tends to 0 rapidly for $\alpha >1$. For $\alpha >1$, long-range correlations develop in time as well as space and the ordered state is similar to that obtained directly from the quantum disordered state discussed in II above.

V. $\alpha \gtrsim 1$: In this case, $y_w$ increases, which forces the flow of $\alpha$  towards 0. But the RG calculation, as is well known, is uncontrolled because the stable fixed point is of the strong-coupling kind. The $z \to \infty, \alpha \to 1^-$ critical point is unstable towards the stable fixed point  $z \to 1, \alpha \to 0$. As mentioned, the transformation of the action in terms of warps and vortices is invalid for $\alpha = 0$. The problem for $\alpha = 0$ is well known to be that of the 3D loop-gas model which has a stable phase which is the same as that of the 3D-XY model. This is also what is found in the quantum Monte-Carlo calculation and represents the ordered phase in Fig. (\ref{fig:phdia}). 

All the principal features of the phase diagram in Fig. (\ref{fig:phdia}) obtained by Monte-Carlo calculations are obtained by the leading order RG calculations above, but some details and some aspects of the correlations are not obtained. The transition from the disordered to the 3D-XY ordered state occurs in the leading RG calculations at $\alpha =1$, while in Monte-Carlo calculations, the ordered phase requires larger $\alpha$ for smaller $JE_c$. Analytic results to obtain such results require higher order RG calculations. The Monte-Carlo calculations reveal that the transition from the quantum-disordered phase to the ordered phase occurs along a line in the $J-\alpha$ plane, whereas the leading order RG results give the transition to be at $\alpha =1$ for all $J < 8\pi$. In the Monte-carlo calculations, it is found also that for fixed $\alpha$, the correlation length $\xi_{\tau}$ varies as $(g-g_c)^{-\nu_{\tau}}$, with $\nu_{\tau} \approx 0.5$, and again with $(\xi_r/a) \propto \log (\xi_{\tau}/\tau_c)$. As noted, Eq.(\ref{warpcorr}) in first order RG, just as in the Monte-carlo calculations, give that $\xi_{\tau}$ has an essential singularity as a function of $(\alpha-\alpha_c)$. This also suggests that the algebraic singularity as a function of $(g-g_c(\alpha_c))$ can only be found in next order RG calculations. It should also be mentioned that the transformation to the topological model relies on a finite dissipation coefficient $\alpha$. One cannot take the limit $\alpha \to 0$ and get the properties of the (2+1) D quantum XY model without dissipation. The passage of the properties of the model from that of the 3D classical XY model at $\alpha = 0$ to those at finite $\alpha$ has been investigated by Monte-Carlo calculations \cite{ZhuHouCMV2016}.

Finally, we recapitulate the results on the correlation functions of the order parameter, which also follow from the RG equations above, and their implication in several experimental problems. Since the action is written in terms of orthogonal variables, the correlation function can be written as a product of correlation of vortices and of warps, as noted earlier.
This results in the remarkable result that the correlation function is separable in space and time. At criticality, the correlation function is 
\be
\label{corfn}
C({\bf r -r}', {\bf \tau-\tau}')\equiv \langle \cos \theta({\bf r}, \tau)\cos \theta({\bf r}', \tau')\rangle \propto \log (|{\bf r - r}'|) e^{-\frac{|{\bf r - r}'|}{\xi_r}} \frac{1}{|\tau - \tau'|}e^{-\frac{|\tau - \tau'|}{\xi_{\tau}}}.
\ee
The Fourier transform gives that the imaginary part of the correlations is 
\be
Im C(\omega, {\bf q}) \propto \frac{1}{\kappa_q^2 + q^2} \tanh\big(\frac{\omega}{\sqrt{4T^2 + \kappa_{\omega}^2}}\big); ~\kappa_q = 1/\xi_{r}, ~\kappa_{\omega} = 1/\xi_{\tau}.
\ee
At criticality, the frequency dependence has precisely the form of the critical fluctuations hypothesized to get the marginal fermi-liquid \cite{CMV-MFL} for fermions scattering from such fluctuations. This together with the separable form of the fluctuations guarantees the linear in $T$ resistivity found in the quantum-critical regime of a variety of quantum-critical metals \cite{CMV-PRL2015}, including the cuprates, some Fe-based antiferromagnetic compounds and some heavy-fermions. This as well as the other quantum-critical properties appear to be unique to scattering from such fluctuations. Direct evidence for fluctuations consistent with such a form has been found in some compounds by neutron scattering \cite{Schroder2, Inosov2010, CMVZhuSchroeder2015}. Such fluctuations have been discovered in the long wave-length limit by Raman scattering in cuprates \cite{Klein_Raman_1991} and deduced over a large region of momentum space as responsible for their strange metal properties as well as promotion of superconductivity \cite{Bok_ScienceADV}. Recently such fluctuations have been used to derive \cite{HouCMV2016} the observed properties of a 2D-XY ferromagnet \cite{AronsonPNAS2014}. \\

Acknowledgements: We gratefully acknowledge very useful discussions with Vivek Aji and H. Krishnamurthy. This work was partially supported by NSF under the grant DMR 1206298.

\bibliographystyle{apsrev4-1}
\bibliography{REF.bib}

\begin{thebibliography}{29}%
\makeatletter
\providecommand \@ifxundefined [1]{%
 \@ifx{#1\undefined}
}%
\providecommand \@ifnum [1]{%
 \ifnum #1\expandafter \@firstoftwo
 \else \expandafter \@secondoftwo
 \fi
}%
\providecommand \@ifx [1]{%
 \ifx #1\expandafter \@firstoftwo
 \else \expandafter \@secondoftwo
 \fi
}%
\providecommand \natexlab [1]{#1}%
\providecommand \enquote  [1]{``#1''}%
\providecommand \bibnamefont  [1]{#1}%
\providecommand \bibfnamefont [1]{#1}%
\providecommand \citenamefont [1]{#1}%
\providecommand \href@noop [0]{\@secondoftwo}%
\providecommand \href [0]{\begingroup \@sanitize@url \@href}%
\providecommand \@href[1]{\@@startlink{#1}\@@href}%
\providecommand \@@href[1]{\endgroup#1\@@endlink}%
\providecommand \@sanitize@url [0]{\catcode `\\12\catcode `\$12\catcode
  `\&12\catcode `\#12\catcode `\^12\catcode `\_12\catcode `\%12\relax}%
\providecommand \@@startlink[1]{}%
\providecommand \@@endlink[0]{}%
\providecommand \url  [0]{\begingroup\@sanitize@url \@url }%
\providecommand \@url [1]{\endgroup\@href {#1}{\urlprefix }}%
\providecommand \urlprefix  [0]{URL }%
\providecommand \Eprint [0]{\href }%
\providecommand \doibase [0]{http://dx.doi.org/}%
\providecommand \selectlanguage [0]{\@gobble}%
\providecommand \bibinfo  [0]{\@secondoftwo}%
\providecommand \bibfield  [0]{\@secondoftwo}%
\providecommand \translation [1]{[#1]}%
\providecommand \BibitemOpen [0]{}%
\providecommand \bibitemStop [0]{}%
\providecommand \bibitemNoStop [0]{.\EOS\space}%
\providecommand \EOS [0]{\spacefactor3000\relax}%
\providecommand \BibitemShut  [1]{\csname bibitem#1\endcsname}%
\let\auto@bib@innerbib\@empty
\bibitem [{\citenamefont {Chakravarty}\ \emph {et~al.}(1988)\citenamefont
  {Chakravarty}, \citenamefont {Ingold}, \citenamefont {S.~Kivelson},\ and\
  \citenamefont {Zimanyi}}]{ChakraKivel1}%
  \BibitemOpen
  \bibfield  {author} {\bibinfo {author} {\bibfnamefont {S.}~\bibnamefont
  {Chakravarty}}, \bibinfo {author} {\bibfnamefont {G.}~\bibnamefont {Ingold}},
  \bibinfo {author} {\bibfnamefont {S.}~\bibnamefont {S.~Kivelson}}, \ and\
  \bibinfo {author} {\bibfnamefont {G.}~\bibnamefont {Zimanyi}},\ }\href@noop
  {} {\bibfield  {journal} {\bibinfo  {journal} {Phys. Rev. B}\ }\textbf
  {\bibinfo {volume} {37}},\ \bibinfo {pages} {3283} (\bibinfo {year}
  {1988})}\BibitemShut {NoStop}%
\bibitem [{\citenamefont {Lin}\ and\ \citenamefont
  {Goldman}(2015)}]{GoldmanRev2014}%
  \BibitemOpen
  \bibfield  {author} {\bibinfo {author} {\bibfnamefont {N.~J.}\ \bibnamefont
  {Lin}, \bibfnamefont {Yen-Hsiang}}\ and\ \bibinfo {author} {\bibfnamefont
  {A.}~\bibnamefont {Goldman}},\ }\href@noop {} {\bibfield  {journal} {\bibinfo
   {journal} {Physica C}\ }\textbf {\bibinfo {volume} {514}},\ \bibinfo {pages}
  {130} (\bibinfo {year} {2015})}\BibitemShut {NoStop}%
\bibitem [{\citenamefont {Wua}\ \emph {et~al.}(2014)\citenamefont {Wua},
  \citenamefont {Kima}, \citenamefont {Parka}, \citenamefont {Tsvelik},\ and\
  \citenamefont {Aronson}}]{AronsonPNAS2014}%
  \BibitemOpen
  \bibfield  {author} {\bibinfo {author} {\bibfnamefont {L.}~\bibnamefont
  {Wua}}, \bibinfo {author} {\bibfnamefont {M.}~\bibnamefont {Kima}}, \bibinfo
  {author} {\bibfnamefont {K.}~\bibnamefont {Parka}}, \bibinfo {author}
  {\bibfnamefont {A.}~\bibnamefont {Tsvelik}}, \ and\ \bibinfo {author}
  {\bibfnamefont {M.}~\bibnamefont {Aronson}},\ }\href@noop {} {\bibfield
  {journal} {\bibinfo  {journal} {PNAS}\ }\textbf {\bibinfo {volume} {39}},\
  \bibinfo {pages} {14088} (\bibinfo {year} {2014})}\BibitemShut {NoStop}%
\bibitem [{\citenamefont {Hou}\ and\ \citenamefont {Varma}(2016)}]{HouCMV2016}%
  \BibitemOpen
  \bibfield  {author} {\bibinfo {author} {\bibfnamefont {C.}~\bibnamefont
  {Hou}}\ and\ \bibinfo {author} {\bibfnamefont {C.~M.}\ \bibnamefont
  {Varma}},\ }\href@noop {} {\  (\bibinfo {year} {2016})}\BibitemShut {NoStop}%
\bibitem [{\citenamefont {Varma}(2015)}]{CMV-PRL2015}%
  \BibitemOpen
  \bibfield  {author} {\bibinfo {author} {\bibfnamefont {C.~M.}\ \bibnamefont
  {Varma}},\ }\href@noop {} {\bibfield  {journal} {\bibinfo  {journal} {Phys.
  Rev. Lett.}\ } (\bibinfo {year} {2015})}\BibitemShut {NoStop}%
\bibitem [{\citenamefont {Varma}(2006)}]{CMV-PRB2006}%
  \BibitemOpen
  \bibfield  {author} {\bibinfo {author} {\bibfnamefont {C.~M.}\ \bibnamefont
  {Varma}},\ }\href@noop {} {\bibfield  {journal} {\bibinfo  {journal} {Phys.
  Rev. B}\ }\textbf {\bibinfo {volume} {73}},\ \bibinfo {pages} {155113}
  (\bibinfo {year} {2006})}\BibitemShut {NoStop}%
\bibitem [{\citenamefont {Kosterlitz}\ and\ \citenamefont
  {Thouless}(1973)}]{KT1973}%
  \BibitemOpen
  \bibfield  {author} {\bibinfo {author} {\bibfnamefont {J.}~\bibnamefont
  {Kosterlitz}}\ and\ \bibinfo {author} {\bibfnamefont {D.}~\bibnamefont
  {Thouless}},\ }\href@noop {} {\bibfield  {journal} {\bibinfo  {journal} {J.
  Phys. C}\ }\textbf {\bibinfo {volume} {6}} (\bibinfo {year}
  {1973})}\BibitemShut {NoStop}%
\bibitem [{\citenamefont {Berezinskii}(1970)}]{Berezinskii}%
  \BibitemOpen
  \bibfield  {author} {\bibinfo {author} {\bibfnamefont {V.}~\bibnamefont
  {Berezinskii}},\ }\href@noop {} {\bibfield  {journal} {\bibinfo  {journal}
  {Zh. Eksp. Teor. Fiz.}\ }\textbf {\bibinfo {volume} {59}},\ \bibinfo {pages}
  {907} (\bibinfo {year} {1970})}\BibitemShut {NoStop}%
\bibitem [{\citenamefont {Stiansen}\ \emph {et~al.}(2012)\citenamefont
  {Stiansen}, \citenamefont {Sperstad},\ and\ \citenamefont
  {Sudb\o{}}}]{Stiansen-PRB2012}%
  \BibitemOpen
  \bibfield  {author} {\bibinfo {author} {\bibfnamefont {E.~B.}\ \bibnamefont
  {Stiansen}}, \bibinfo {author} {\bibfnamefont {I.~B.}\ \bibnamefont
  {Sperstad}}, \ and\ \bibinfo {author} {\bibfnamefont {A.}~\bibnamefont
  {Sudb\o{}}},\ }\href {\doibase 10.1103/PhysRevB.85.224531} {\bibfield
  {journal} {\bibinfo  {journal} {Phys. Rev. B}\ }\textbf {\bibinfo {volume}
  {85}},\ \bibinfo {pages} {224531} (\bibinfo {year} {2012})}\BibitemShut
  {NoStop}%
\bibitem [{\citenamefont {Zhu}\ \emph {et~al.}(2015)\citenamefont {Zhu},
  \citenamefont {Chen},\ and\ \citenamefont {Varma}}]{ZhuChenCMV2015}%
  \BibitemOpen
  \bibfield  {author} {\bibinfo {author} {\bibfnamefont {L.}~\bibnamefont
  {Zhu}}, \bibinfo {author} {\bibfnamefont {Y.}~\bibnamefont {Chen}}, \ and\
  \bibinfo {author} {\bibfnamefont {C.~M.}\ \bibnamefont {Varma}},\ }\href
  {\doibase 10.1103/PhysRevB.91.205129} {\bibfield  {journal} {\bibinfo
  {journal} {Phys. Rev. B}\ }\textbf {\bibinfo {volume} {91}},\ \bibinfo
  {pages} {205129} (\bibinfo {year} {2015})}\BibitemShut {NoStop}%
\bibitem [{\citenamefont {Aji}\ and\ \citenamefont {Varma}(2007)}]{Aji-V-qcf1}%
  \BibitemOpen
  \bibfield  {author} {\bibinfo {author} {\bibfnamefont {V.}~\bibnamefont
  {Aji}}\ and\ \bibinfo {author} {\bibfnamefont {C.~M.}\ \bibnamefont
  {Varma}},\ }\href {\doibase 10.1103/PhysRevLett.99.067003} {\bibfield
  {journal} {\bibinfo  {journal} {Phys. Rev. Lett.}\ }\textbf {\bibinfo
  {volume} {99}},\ \bibinfo {pages} {067003} (\bibinfo {year}
  {2007})}\BibitemShut {NoStop}%
\bibitem [{\citenamefont {Aji}\ and\ \citenamefont {Varma}(2010)}]{Aji-V-qcf3}%
  \BibitemOpen
  \bibfield  {author} {\bibinfo {author} {\bibfnamefont {V.}~\bibnamefont
  {Aji}}\ and\ \bibinfo {author} {\bibfnamefont {C.~M.}\ \bibnamefont
  {Varma}},\ }\href {\doibase 10.1103/PhysRevB.82.174501} {\bibfield  {journal}
  {\bibinfo  {journal} {Phys. Rev. B}\ }\textbf {\bibinfo {volume} {82}},\
  \bibinfo {pages} {174501} (\bibinfo {year} {2010})}\BibitemShut {NoStop}%
\bibitem [{\citenamefont {Moriya}(1985)}]{Moriya-book}%
  \BibitemOpen
  \bibfield  {author} {\bibinfo {author} {\bibfnamefont {T.}~\bibnamefont
  {Moriya}},\ }\href@noop {} {\emph {\bibinfo {title} {Spin Fluctuations in
  Itinerant Electron Magnetism}}}\ (\bibinfo  {publisher} {Springer-Verlag,
  Berlin},\ \bibinfo {year} {1985})\BibitemShut {NoStop}%
\bibitem [{\citenamefont {Hertz}(1976)}]{Hertz}%
  \BibitemOpen
  \bibfield  {author} {\bibinfo {author} {\bibfnamefont {J.~A.}\ \bibnamefont
  {Hertz}},\ }\href@noop {} {\bibfield  {journal} {\bibinfo  {journal} {Phys.
  Rev. B}\ }\textbf {\bibinfo {volume} {14}},\ \bibinfo {pages} {1165}
  (\bibinfo {year} {1976})}\BibitemShut {NoStop}%
\bibitem [{\citenamefont {Caldeira}\ and\ \citenamefont
  {Leggett}(1983)}]{CaldeiraLeggett}%
  \BibitemOpen
  \bibfield  {author} {\bibinfo {author} {\bibfnamefont {A.}~\bibnamefont
  {Caldeira}}\ and\ \bibinfo {author} {\bibfnamefont {A.}~\bibnamefont
  {Leggett}},\ }\href@noop {} {\bibfield  {journal} {\bibinfo  {journal} {Ann.
  Phys. (NY)}\ }\textbf {\bibinfo {volume} {149}},\ \bibinfo {pages} {374}
  (\bibinfo {year} {1983})}\BibitemShut {NoStop}%
\bibitem [{\citenamefont {Zhu}\ \emph {et~al.}()\citenamefont {Zhu},
  \citenamefont {Hou},\ and\ \citenamefont {Varma}}]{ZhuHouCMV2016}%
  \BibitemOpen
  \bibfield  {author} {\bibinfo {author} {\bibfnamefont {L.}~\bibnamefont
  {Zhu}}, \bibinfo {author} {\bibfnamefont {C.}~\bibnamefont {Hou}}, \ and\
  \bibinfo {author} {\bibfnamefont {C.~M.}\ \bibnamefont {Varma}},\ }\href@noop
  {} {\ }\BibitemShut {NoStop}%
\bibitem [{\citenamefont {Villain}(1975)}]{Villain}%
  \BibitemOpen
  \bibfield  {author} {\bibinfo {author} {\bibfnamefont {J.}~\bibnamefont
  {Villain}},\ }\href@noop {} {\bibfield  {journal} {\bibinfo  {journal} {J.
  Phys. (Paris)}\ }\textbf {\bibinfo {volume} {36}},\ \bibinfo {pages} {581}
  (\bibinfo {year} {1975})}\BibitemShut {NoStop}%
\bibitem [{\citenamefont {Polyakov}(1977)}]{polyakov}%
  \BibitemOpen
  \bibfield  {author} {\bibinfo {author} {\bibfnamefont {A.}~\bibnamefont
  {Polyakov}},\ }\href@noop {} {\bibfield  {journal} {\bibinfo  {journal}
  {Nucl. Phys. B}\ }\textbf {\bibinfo {volume} {120}},\ \bibinfo {pages} {429}
  (\bibinfo {year} {1977})}\BibitemShut {NoStop}%
\bibitem [{\citenamefont {Kosterlitz}(1974)}]{Kosterlitz1974}%
  \BibitemOpen
  \bibfield  {author} {\bibinfo {author} {\bibfnamefont {J.}~\bibnamefont
  {Kosterlitz}},\ }\href@noop {} {\bibfield  {journal} {\bibinfo  {journal} {J.
  Phys. C}\ }\textbf {\bibinfo {volume} {7}},\ \bibinfo {pages} {1046}
  (\bibinfo {year} {1974})}\BibitemShut {NoStop}%
\bibitem [{\citenamefont {Jose}\ \emph {et~al.}(1977)\citenamefont {Jose},
  \citenamefont {L.~P.~Kadanoff}, \citenamefont {Kirkpatrick},\ and\
  \citenamefont {Nelson}}]{JKKN1977}%
  \BibitemOpen
  \bibfield  {author} {\bibinfo {author} {\bibfnamefont {J.}~\bibnamefont
  {Jose}}, \bibinfo {author} {\bibfnamefont {L.}~\bibnamefont
  {L.~P.~Kadanoff}}, \bibinfo {author} {\bibfnamefont {S.}~\bibnamefont
  {Kirkpatrick}}, \ and\ \bibinfo {author} {\bibfnamefont {D.}~\bibnamefont
  {Nelson}},\ }\href@noop {} {\bibfield  {journal} {\bibinfo  {journal} {Phys.
  Rev. B}\ }\textbf {\bibinfo {volume} {16}},\ \bibinfo {pages} {1217}
  (\bibinfo {year} {1977})}\BibitemShut {NoStop}%
\bibitem [{\citenamefont {Kosterlitz}(1977)}]{Kosterlitz1977}%
  \BibitemOpen
  \bibfield  {author} {\bibinfo {author} {\bibfnamefont {J.~M.}\ \bibnamefont
  {Kosterlitz}},\ }\href@noop {} {\bibfield  {journal} {\bibinfo  {journal} {J.
  Phys. C}\ }\textbf {\bibinfo {volume} {10}},\ \bibinfo {pages} {3753}
  (\bibinfo {year} {1977})}\BibitemShut {NoStop}%
\bibitem [{\citenamefont {Aji}\ and\ \citenamefont {Varma}(2009)}]{Aji-V-qcf2}%
  \BibitemOpen
  \bibfield  {author} {\bibinfo {author} {\bibfnamefont {V.}~\bibnamefont
  {Aji}}\ and\ \bibinfo {author} {\bibfnamefont {C.~M.}\ \bibnamefont
  {Varma}},\ }\href {\doibase 10.1103/PhysRevB.79.184501} {\bibfield  {journal}
  {\bibinfo  {journal} {Phys. Rev. B}\ }\textbf {\bibinfo {volume} {79}},\
  \bibinfo {pages} {184501} (\bibinfo {year} {2009})}\BibitemShut {NoStop}%
\bibitem [{\citenamefont {Breznay}\ \emph {et~al.}(2015)\citenamefont
  {Breznay}, \citenamefont {Steiner}, \citenamefont {Kivelson},\ and\
  \citenamefont {Kapitulnik}}]{KapitulnikSIT1}%
  \BibitemOpen
  \bibfield  {author} {\bibinfo {author} {\bibfnamefont {N.}~\bibnamefont
  {Breznay}}, \bibinfo {author} {\bibfnamefont {M.}~\bibnamefont {Steiner}},
  \bibinfo {author} {\bibfnamefont {S.~A.}\ \bibnamefont {Kivelson}}, \ and\
  \bibinfo {author} {\bibfnamefont {A.}~\bibnamefont {Kapitulnik}},\
  }\href@noop {} {\bibfield  {journal} {\bibinfo  {journal} {PNAS}\ }\textbf
  {\bibinfo {volume} {6}},\ \bibinfo {pages} {1} (\bibinfo {year}
  {2015})}\BibitemShut {NoStop}%
\bibitem [{\citenamefont {Varma}\ \emph {et~al.}(1989)\citenamefont {Varma},
  \citenamefont {Littlewood}, \citenamefont {Schmitt-Rink}, \citenamefont
  {Abrahams},\ and\ \citenamefont {Ruckenstein}}]{CMV-MFL}%
  \BibitemOpen
  \bibfield  {author} {\bibinfo {author} {\bibfnamefont {C.~M.}\ \bibnamefont
  {Varma}}, \bibinfo {author} {\bibfnamefont {P.~B.}\ \bibnamefont
  {Littlewood}}, \bibinfo {author} {\bibfnamefont {S.}~\bibnamefont
  {Schmitt-Rink}}, \bibinfo {author} {\bibfnamefont {E.}~\bibnamefont
  {Abrahams}}, \ and\ \bibinfo {author} {\bibfnamefont {A.~E.}\ \bibnamefont
  {Ruckenstein}},\ }\href {\doibase 10.1103/PhysRevLett.63.1996} {\bibfield
  {journal} {\bibinfo  {journal} {Phys. Rev. Lett.}\ }\textbf {\bibinfo
  {volume} {63}},\ \bibinfo {pages} {1996} (\bibinfo {year}
  {1989})}\BibitemShut {NoStop}%
\bibitem [{\citenamefont {Schr\"oder}\ and\ \citenamefont
  {et~al.}(2000)}]{Schroder2}%
  \BibitemOpen
  \bibfield  {author} {\bibinfo {author} {\bibfnamefont {A.}~\bibnamefont
  {Schr\"oder}}\ and\ \bibinfo {author} {\bibnamefont {et~al.}},\ }\href@noop
  {} {\bibfield  {journal} {\bibinfo  {journal} {Nature (London)}\ }\textbf
  {\bibinfo {volume} {407}},\ \bibinfo {pages} {351} (\bibinfo {year}
  {2000})}\BibitemShut {NoStop}%
\bibitem [{\citenamefont {Inosov}\ \emph {et~al.}(2010)\citenamefont {Inosov},
  \citenamefont {Park}, \citenamefont {Bourges}, \citenamefont {Sun},
  \citenamefont {Sidis}, \citenamefont {Schneidewind}, \citenamefont {Hradil},
  \citenamefont {Haug}, \citenamefont {Lin}, \citenamefont {Keimer},\ and\
  \citenamefont {Hinkov}}]{Inosov2010}%
  \BibitemOpen
  \bibfield  {author} {\bibinfo {author} {\bibfnamefont {D.~S.}\ \bibnamefont
  {Inosov}}, \bibinfo {author} {\bibfnamefont {J.~T.}\ \bibnamefont {Park}},
  \bibinfo {author} {\bibfnamefont {P.}~\bibnamefont {Bourges}}, \bibinfo
  {author} {\bibfnamefont {D.~L.}\ \bibnamefont {Sun}}, \bibinfo {author}
  {\bibfnamefont {Y.}~\bibnamefont {Sidis}}, \bibinfo {author} {\bibfnamefont
  {A.}~\bibnamefont {Schneidewind}}, \bibinfo {author} {\bibfnamefont
  {K.}~\bibnamefont {Hradil}}, \bibinfo {author} {\bibfnamefont
  {D.}~\bibnamefont {Haug}}, \bibinfo {author} {\bibfnamefont {C.~T.}\
  \bibnamefont {Lin}}, \bibinfo {author} {\bibfnamefont {B.}~\bibnamefont
  {Keimer}}, \ and\ \bibinfo {author} {\bibfnamefont {V.}~\bibnamefont
  {Hinkov}},\ }\href@noop {} {\bibfield  {journal} {\bibinfo  {journal} {Nature
  Phys.}\ }\textbf {\bibinfo {volume} {6}},\ \bibinfo {pages} {178} (\bibinfo
  {year} {2010})}\BibitemShut {NoStop}%
\bibitem [{\citenamefont {Varma}\ \emph {et~al.}(2015)\citenamefont {Varma},
  \citenamefont {Zhu},\ and\ \citenamefont {Schroeder}}]{CMVZhuSchroeder2015}%
  \BibitemOpen
  \bibfield  {author} {\bibinfo {author} {\bibfnamefont {C.~M.}\ \bibnamefont
  {Varma}}, \bibinfo {author} {\bibfnamefont {L.}~\bibnamefont {Zhu}}, \ and\
  \bibinfo {author} {\bibfnamefont {A.}~\bibnamefont {Schroeder}},\ }\href@noop
  {} {\bibfield  {journal} {\bibinfo  {journal} {Phys. Rev. B}\ }\textbf
  {\bibinfo {volume} {92}},\ \bibinfo {pages} {155150} (\bibinfo {year}
  {2015})}\BibitemShut {NoStop}%
\bibitem [{\citenamefont {Slakey}\ \emph {et~al.}(1991)\citenamefont {Slakey},
  \citenamefont {Klein}, \citenamefont {Rice},\ and\ \citenamefont
  {Ginsberg}}]{Klein_Raman_1991}%
  \BibitemOpen
  \bibfield  {author} {\bibinfo {author} {\bibfnamefont {F.}~\bibnamefont
  {Slakey}}, \bibinfo {author} {\bibfnamefont {M.~V.}\ \bibnamefont {Klein}},
  \bibinfo {author} {\bibfnamefont {J.~P.}\ \bibnamefont {Rice}}, \ and\
  \bibinfo {author} {\bibfnamefont {D.~M.}\ \bibnamefont {Ginsberg}},\
  }\href@noop {} {\bibfield  {journal} {\bibinfo  {journal} {Phys. Rev. B}\
  }\textbf {\bibinfo {volume} {43}},\ \bibinfo {pages} {3764(R)} (\bibinfo
  {year} {1991})}\BibitemShut {NoStop}%
\bibitem [{\citenamefont {Bok}\ and\ \citenamefont
  {et~al.}(2016)}]{Bok_ScienceADV}%
  \BibitemOpen
  \bibfield  {author} {\bibinfo {author} {\bibfnamefont {J.~M.}\ \bibnamefont
  {Bok}}\ and\ \bibinfo {author} {\bibnamefont {et~al.}},\ }\href@noop {}
  {\bibfield  {journal} {\bibinfo  {journal} {Science Advances}\ }\textbf
  {\bibinfo {volume} {2}},\ \bibinfo {pages} {e1501329} (\bibinfo {year}
  {2016})}\BibitemShut {NoStop}%
\end{thebibliography}%
\newpage
{\bf Supplementary Information}\\
\section{Derivation of the Renormalization Group Equations for Warps}
Since warps and vortices are orthogonal objects, one may consider the partition function as the product of their partition functions.
As in Eq. (5) of the text, the action for the warps, $S_w = S_w^0 + S_w^{\prime}$ consists of two terms, 
\be
\label{Swarp}
S_w=\alpha\sum_{i\neq j}\rho_w(\bm{r}_i,\tau_i)\ln\frac{|\tau_i-\tau_j|}{\tau_c}\rho_w(\bm{r}_i,\tau_j) \delta(\bm{r}_i-\bm{r}_j)\\ 
S_w^{\prime}=-\frac{\sqrt{JC}}{4\pi}\sum_{i\neq j}\rho_w(\bm{r}_i,\tau_i)\frac{1}{\sqrt{v^2(\tau_i-\tau_j)^2+|\bm{r}_i-\bm{r}_j|^2}}\rho_w(\bm{r}_j,\tau_j)
\ee
where $v=\sqrt{\frac{Jc^2}{C}}$. This must be supplemented by terms which take into account the short distance or the core-energy of the topological defects. 

The partition function for warps is
\be
\label{partfn}
Z &=& \sum_ny_w^{2n}\int_{a_c}^{\infty}\frac{d^2r_{2n}}{a_c^2}\cdots\int_{a_c}^{+\infty}\frac{d^2r_1}{a_c^2}\int_0^{\beta}\frac{d\tau_{2n}}{\tau_c}\cdots\int_0^{\tau_2-\tau_c}\frac{d\tau_1}{\tau_c}\sum_{\{\rho_w=\pm1\}}\mathrm{exp}\bigg[\alpha\sum_{i\neq j}\int d^2r_i\rho_w(\bm{r}_i,\tau_i)\ln\frac{|\tau_i-\tau_j|}{\tau_c}\delta(\bm{r}_i-\bm{r}_j)\rho_w(\bm{r}_j,\tau_j)\nonumber\\&+&g\sum_{i\neq j}\rho_w(\bm{r}_i,\tau_i)\frac{1}{\sqrt{v^2(\tau_i-\tau_j)^2+|\bm{r}_i-\bm{r}_j|^2}}\rho_w(\bm{r}_j,\tau_j)\bigg].
\ee
Here we have defined $y_w=ya_c^2\tau_c$, $y$ is the fugacity of the warps, and $g=\frac{\sqrt{JE_c}}{4\pi}$, and we have normalized the space and time integrals to dimensionless variables in terms of the lattice constant $a$ and the upper cut-off in time, $\tau_c$.
We consider only $\rho_w(r,\tau)=\pm 1$, as higher charged warps are unimportant for low energy phenomena. 

The first term in the partition function is much more singular than the second term. In the first term warps interact locally in space. We consider renormalization of interactions between a pair of warps due to a pair of virtual warps by summing over all possible interactions between the virtual pair and the others.  The renormalized interactions at longer and longer distances and longer and longer times are found by integrating over spatial scales increasing by $e^{d\ell_r}$ and time-scales increasing by $e^{d\ell_{\tau}}$. Let the warps be located at $(\bm{r}_i,\tau_i)$ and $(\bm{r}_j,\tau_j)$ with charge $\rho_w(\bm{r}_i,\tau_i)=+1$ and $\rho_w(\bm{r}_j,\tau_j)=-1$, respectively. The virtual warps pair have $\rho_w(\bm{r}^\prime,\tau^\prime)=+1$ and $\rho_w(\bm{r}^{\prime\prime},\tau^{\prime\prime})=-1$ and satisfy $a_c<|\bm{r}^\prime-\bm{r}^{\prime\prime}|<a_ce^{dl_r}$ and $\tau_c<|\tau^\prime-\tau^{\prime\prime}|<\tau_c e^{dl_\tau}$. Let $r_i=(\bm{r}_i,\tau_i)$ and define,
\be
U(r_i,r_j)=\alpha\delta(\bm{r}_i-\bm{r}_j)\ln\frac{|\tau_i-\tau_j|}{\tau_c}+g\frac{1}{\sqrt{v^2(\tau_i-\tau_j)^2+|\bm{r}_i-\bm{r}_j|^2}},
\ee
The effective interaction of the pair of warps is given by
\be
e^{-U_{\text{eff}}(r_i,r_j)}&=&<e^{-U(r_i,r_j)}>|_{short}=e^{-U(r_i,r_j)-\delta U(r_i,r_j)},
\ee
where the expectation value is the statistical average over the partition function in Eq.(\ref{partfn}) over the integrated short scale. In the second equality, the effective interaction is written as the bare interaction plus the renormalized interaction after integrating the virtual pair. To lowest order in $y_w$, we have  
\be
\label{deltaU}
e^{-\delta U(r_i,r_j)}&=&\frac{1+y_w^2\int\frac{d^2r^\prime}{a_c^2}\frac{d^2r^{\prime\prime}}{a_c^2}\int\frac{d\tau^\prime}{\tau_c}\int\frac{d\tau^{\prime\prime}}{\tau_c}e^{-U(r^\prime,r^{\prime\prime})}\mathrm{e}^{C(r_i,r_i;r^\prime,r^{\prime\prime})+D(r_i,r_j;r^\prime,r^{\prime\prime})}+O(y_w^4)}{1+y_w^2\int\frac{d^2r^\prime}{a_c^2}\frac{d^2r^{\prime\prime}}{a_c^2}\int\frac{d\tau^\prime}{\tau_c}\int\frac{d\tau^{\prime\prime}}{\tau_c}e^{-U(r^\prime,r^{\prime\prime})}+O(y_w^4)}\nonumber\\
&=&1+y_w^2\int\frac{d^2r^\prime}{a_c}\frac{d^2r^{\prime\prime}}{a_c}\int\frac{d\tau^\prime}{\tau_c}\int\frac{d\tau^{\prime\prime}}{\tau_c}e^{-U(r^\prime,r^{\prime\prime})}\bigg[e^{C(r_i,r_i;r^\prime,r^{\prime\prime})+D(r_i,r_j;r^\prime,r^{\prime\prime})}-1\bigg]+O(y_w^4),\nonumber\\
\ee
where $C(r_i,r_i;r^\prime,r^{\prime\prime})$ is the contribution from virtual pairs to the interaction of two warps at the same space site, $D(r_i,r_j;r^\prime,r^{\prime\prime})$is the contribution from virtual pairs to the interaction of two warps at a different time and space site. We divide the term $C$  into five parts: (I) $\bm{r}_i=\bm{r}^\prime=\bm{r}^{\prime\prime}$; (II) $\bm{r}_i\neq\bm{r}^\prime,\bm{r}^\prime=\bm{r}^{\prime\prime}$; (III) $\bm{r}_i\neq\bm{r}^\prime\neq\bm{r}^{\prime\prime}$; (IV) $\bm{r}_i=\bm{r}^\prime,\bm{r}^\prime\neq\bm{r}^{\prime\prime}$; (V) $\bm{r}_i=\bm{r}^{\prime\prime},\bm{r}^\prime\neq\bm{r}^{\prime\prime}$. By summing over the two charge distribution of neutral virtual pair, only the first two terms  are non-zero. We then have
\be
C_1&=&\bigg[\alpha\ln\frac{|\tau_i-\tau^\prime|}{|\tau_i-\tau^{\prime\prime}|}\frac{|\tau_j-\tau^{\prime\prime}|}{|\tau_j-\tau^\prime|}\bigg]\delta(\bm{r}^\prime-\bm{r}^{\prime\prime})\delta(\bm{r}_i-\bm{r}^\prime),\nonumber\\
C_2&=&\frac{g}{\sqrt{(\bm{r}_i-\bm{r}^\prime)^2+v^2(\tau_i-\tau^\prime)^2}}-\frac{g}{\sqrt{(\bm{r}_i-\bm{r}^{\prime\prime})^2+v^2(\tau_i-\tau^{\prime\prime})^2}}-\frac{g}{\sqrt{(\bm{r}_i-\bm{r}^{\prime})^2+v^2(\tau_j-\tau^{\prime})^2}}\nonumber\\&&+\frac{g}{\sqrt{(\bm{r}_i-\bm{r}^{\prime\prime})^2+v^2(\tau_j-\tau^{\prime\prime})^2}}.
\ee
For $D$ term, we divide the whole space and time into nine piece: (I) $\bm{r}_i\neq\bm{r}^\prime\neq\bm{r}^{\prime\prime}\neq\bm{r}_j$; (II) $\bm{r}_i=\bm{r}^\prime\neq\bm{r}^{\prime\prime}\neq\bm{r}_j$; (III) $\bm{r}_i=\bm{r}^{\prime\prime}\neq\bm{r}^{\prime}\neq\bm{r}_j$; (IV) $\bm{r}_j=\bm{r}^{\prime}\neq\bm{r}^{\prime\prime}\neq\bm{r}_i$; (V) $\bm{r}_j=\bm{r}^{\prime\prime}\neq\bm{r}^{\prime}\neq\bm{r}_i$; (VI) $\bm{r}_i=\bm{r}^{\prime}\neq\bm{r}^{\prime\prime}=\bm{r}_j$; (VII) $\bm{r}_i=\bm{r}^{\prime\prime}\neq\bm{r}^{\prime}=\bm{r}_j$; (VIII) $\bm{r}_i=\bm{r}^{\prime}=\bm{r}^{\prime\prime}\neq\bm{r}_j$; (IX) $\bm{r}_j=\bm{r}^{\prime}=\bm{r}^{\prime\prime}\neq\bm{r}_i$. Summing over the two charge distribution of  the neutral pair, only the first term is non-zero:
\be
D_1=\frac{g}{\sqrt{(\bm{r}_i-\bm{r}^\prime)^2+v^2(\tau_i-\tau^\prime)^2}}-\frac{g}{\sqrt{(\bm{r}_i-\bm{r}^{\prime\prime})^2+v^2(\tau_i-\tau^{\prime\prime})^2}}-\frac{g}{\sqrt{(\bm{r}_j-\bm{r}^{\prime})^2+v^2(\tau_j-\tau^{\prime})^2}}+\frac{g}{\sqrt{(\bm{r}_j-\bm{r}^{\prime\prime})^2+v^2(\tau_j-\tau^{\prime\prime})^2}}.\nonumber\\
\ee
Now we evaluate (\ref{deltaU}). For the contribution from $C$, we find the term proportion to $\ln\frac{\tau_j-\tau_i}{\tau_c}$, and for the contribution from $D$,  we will find the term proportion to $g/r$. Let us assume $\tau_i<\tau^{\prime\prime}<\tau^\prime<\tau_j$. Then
\be
\int\frac{d^2r^\prime}{a_c^2}\frac{d^2r^{\prime\prime}}{a_c^2}\int\frac{d\tau^\prime}{\tau_c}\frac{d\tau^{\prime\prime}}{\tau_c}C_1&=&\frac{dl_\tau}{a_c^4}\bigg[2\alpha\ln\frac{\tau_j-\tau_i}{\tau_c}\bigg],\nonumber\\
\int\frac{d^2r^\prime}{a_c^2}\frac{d^2r^{\prime\prime}}{a_c^2}\int\frac{d\tau^\prime}{\tau_c}\frac{d\tau^{\prime\prime}}{\tau_c}C_2&=&-4\pi^2(dl_\tau+dl_r)\frac{g}{a_c^2}(\tau_j-\tau_i).
\ee
and
\be
\int\frac{d^2r^\prime}{a_c^2}\frac{d^2r^{\prime\prime}}{a_c^2}\int\frac{d\tau^\prime}{\tau_c}\frac{d\tau^{\prime\prime}}{\tau_c}D_1=-4\pi^2(dl_\tau+dl_r)\frac{g}{a_c^2}(\tau_j-\tau_i).
\ee
So, we have
\be
e^{-\delta U(r_i,r_j)}&=&1+y_w^2\int\frac{d^2r^\prime}{a_c}\frac{d^2r^{\prime\prime}}{a_c}\int\frac{d\tau^\prime}{\tau_c}\int\frac{d\tau^{\prime\prime}}{\tau_c}\bigg[C_1+\frac{1}{2}D_1^2\bigg]\nonumber\\
&=&1+2\alpha\frac{y_w^2}{a_c^4}dl_\tau\ln\frac{\tau_j-\tau_i}{\tau_c}+\frac{1}{2}y_w^2\int\frac{d^2r^\prime}{a_c}\frac{d^2r^{\prime\prime}}{a_c}\int\frac{d\tau^\prime}{\tau_c}\int\frac{d\tau^{\prime\prime}}{\tau_c}D_1^2.
\ee
Let
\be
K&=&\frac{1}{2}\int\frac{d^2r^\prime}{a_c^2}\frac{d^2r^{\prime\prime}}{a_c^2}\int\frac{d\tau^\prime}{\tau_c}\int\frac{d\tau^{\prime\prime}}{\tau_c}D_1^2\nonumber\\
&=&\frac{1}{2}g^2\int\frac{d^2R}{a_c^2}\int\frac{d^2r}{a_c^2}\int\frac{d\tau_s}{\tau_c}\int\frac{d\tau}{\tau_c}\bigg[\frac{[\bm{r}\cdot(\bm{r}_i-\bm{R})+v^2\tau(\tau_i-\tau_s)]^2}{[(\bm{r}_i-\bm{R})^2+v^2(\tau_i-\tau_s)^2]^{3}}\nonumber\\&+&\frac{[\bm{r}\cdot(\bm{r}_j-\bm{R})+v^2\tau(\tau_j-\tau_s)]^2}{[(\bm{r}_j-\bm{R})^2+v^2(\tau_j-\tau_s)^2]^{3}}-2\frac{\bm{r}\cdot(\bm{r}_i-\bm{R})+v^2\tau(\tau_i-\tau_s)}{[(\bm{r}_i-\bm{R})^2+v^2(\tau_i-\tau_s)^2]^{3/2}}\frac{\bm{r}\cdot(\bm{r}_j-\bm{R})+v^2\tau(\tau_j-\tau_s)}{[(\bm{r}_j-\bm{R})^2+v^2(\tau_j-\tau_s)^2]^{3/2}}\bigg]. \nonumber\\
\ee
where $\tau=\tau^\prime-\tau^{\prime\prime}$ and $\tau_s=(\tau^\prime+\tau^{\prime\prime})/2$.  By integrating out the coordinates of center of mass of the virtual pair and over longer spatial and time scales, and keeping terms of $O(dl_r)$ and $O(dl_\tau)$, we arrive at
\be
K=-\frac{8\pi^3}{3}\frac{g^2}{v\tau_c}\bigg[(\frac{1}{4}+\frac{v^2}{2v_c^2})dl_\tau+(1+\frac{v^2}{3v_c^2})dl_r\bigg]\frac{1}{\sqrt{(\bm{r}_i-\bm{r}_j)^2+v^2(\tau_i-\tau_j)^2}}.
\ee
Adding all contributions, we get
\be
e^{-\delta U(r_i,r_j)}=1&+&y_w^2\bigg\{2\alpha\frac{dl_\tau}{a_c^4}\ln\frac{\tau_j-\tau_i}{\tau_c}-\frac{8\pi^3}{3}\frac{g^2}{v\tau_c}\bigg[(\frac{1}{4}+\frac{v^2}{2v_c^2})dl_\tau+(1+\frac{v^2}{3v_c^2})dl_r\bigg]\nonumber\\&\times&\frac{1}{\sqrt{(\bm{r}_i-\bm{r}_j)^2+v^2(\tau_i-\tau_j)^2}}\bigg\}.
\ee
By re-exponentiation, we get
\be
e^{-U_{\text{eff}}}=e^{-U(r_i,r_j)}e^{\bigg(2y_w^2dl_\tau\frac{\alpha}{a_c^4}\ln\frac{\tau_j-\tau_i}{\tau_c}-\frac{8\pi^3}{3}\frac{g^2}{v\tau_c}[(\frac{1}{4}+\frac{v^2}{2v_c^2})dl_\tau+(1+\frac{v^2}{3v_c^2})dl_r]\frac{1}{\sqrt{(\bm{r}_i-\bm{r}_j)^2+v^2(\tau_i-\tau_j)^2}}\bigg)}\nonumber\\
\ee
We can see that the corrected effective interactions due to renormalization of the length and time-scales are given by,
\be
{\alpha}_{ren}&=&\alpha-2\alpha\frac{y_w^2}{a_c^4}dl_\tau,\nonumber\\
{g}_{ren}&=&g-\frac{8\pi^3}{3}\frac{g^2}{v\tau_c}\bigg[(\frac{1}{4}+\frac{v^2}{2v_c^2})dl_\tau+(1+\frac{v^2}{3v_c^2})dl_r\bigg].
\ee
Now in our renormalized action, the short cutoff becomes $\tau_ce^{dl_\tau}$ and $a_ce^{dl_r}$.
\be
Z_r&=& \sum_ny_w^{2n}\int_{a_ce^{dl_r}}^{+\infty}\frac{d^2r_{2n}}{a_c^2}\cdots\int_{a_ce^{dl_\tau}}\frac{d^2r_1}{a_c^2}\int_0^{\beta}\frac{d\tau_{2n}}{\tau_c}\cdots\int_0^{\tau_2-\tau_ce^{dl_\tau}}\frac{d\tau_1}{\tau_c}\nonumber\\&&\sum_{\{\rho_w=\pm1\}}\mathrm{exp}\bigg[{\alpha}_{ren}\sum_{i\neq j}\int d^2r_i\rho_w(\bm{r}_i,\tau_i)\ln\frac{|\tau_i-\tau_j|}{\tau_c}\delta(\bm{r}_i-\bm{r}_j)\rho_w(\bm{r}_j,\tau_j)+{g}_{ren}\sum_{i\neq j}\rho_w(\bm{r}_i,\tau_i)\frac{1}{\sqrt{v^2(\tau_i-\tau_j)^2+|\bm{r}_i-\bm{r}_j|^2}}\rho_w(\bm{r}_j,\tau_j)\bigg].\nonumber\\
\ee
We need to rescale $\bm{r}\to\bm{r}e^{-dl_r}$ and $\tau\to\tau e^{-dl_\tau}$ to get back the original action. Doing so, we have
\be
Z_r&=& \sum_ny_w^{2n}e^{4ndl_r}e^{2ndl_\tau}\int_{a_c}^{+\infty}\frac{d^2r_{2n}}{a_c^2}\cdots\int_{a_c}\frac{d^2r_1}{a_c^2}\int_0^{\beta}\frac{d\tau_{2n}}{\tau_c}\cdots\int_0^{\tau_2-\tau_c}\frac{d\tau_1}{\tau_c}\nonumber\\&&\sum_{\{\rho_w=\pm1\}}\mathrm{exp}\bigg[{\alpha}_{ren}\sum_{i\neq j}\int d^2r_i\rho_w(\bm{r}_i,\tau_i)\ln\frac{|\tau_i-\tau_j|e^{dl_\tau}}{\tau_c}\delta(\bm{r}_i-\bm{r}_j)\rho_w(\bm{r}_j,\tau_j)\nonumber\\&+&{g}_{ren}\sum_{i\neq j}\rho_w(\bm{r}_i,\tau_i)\frac{e^{-dl_r}}{\sqrt{v^2e^{2dl_\tau-2dl_r}(\tau_i-\tau_j)^2+|\bm{r}_i-\bm{r}_j|^2}}\rho_w(\bm{r}_j,\tau_j)\bigg].\nonumber\\
\ee
This contributes extra correction to $g$, $y_w$, and $v$
\be
dg_{rescale}&=&-gdl_r,\nonumber\\
dy_{w,rescale}&=&\bigg\{(1-\alpha)dl_\tau+(2-g)dl_r\bigg\}y_w,\nonumber\\
dv_{rescale}&=&(dl_\tau-dl_r)v.
\ee
Finally, we arrive the renormalization equations for warps by adding up the contribution from renormalization and rescaling
\be
\label{rgeqs}
d\alpha&=&-2\alpha\frac{y_w^2}{a_c^4}dl_\tau,\nonumber\\
dg&=&-gdl_r-\frac{8\pi^3}{3}\frac{g^2}{v\tau_c}\bigg[(\frac{1}{4}+\frac{v^2}{2v_c^2})dl_\tau+(1+\frac{v^2}{3v_c^2})dl_r\bigg],\nonumber\\
dy_w&=&\bigg[(1-\alpha)dl_\tau+(2-g)dl_r\bigg]y_w,\nonumber\\
dv&=&(dl_\tau-dl_r)v.
\ee
Eqs. (\ref{rgeqs}) are reproduced in the main part of the paper and used to analyze the renormalization group flows.

\end{document}